\documentclass[aps,prl,reprint,showpacs,superscriptaddress,floatfix,twocolumn]{revtex4-1}
\usepackage{color}
\usepackage{ulem}
\usepackage{amsmath}
\usepackage{bbm}
\usepackage[ansinew]{inputenc}
\usepackage[english]{babel}
\usepackage{graphicx}
\usepackage{mathrsfs}

\newcommand{\HBi}{\ensuremath{\text{Bi}^{82+}}}
\newcommand{\LiBi}{\ensuremath{\text{Bi}^{80+}}}

\begin{document}
\hspace{5.2in} \mbox{Fermilab-Pub-04/xxx-E}

\title{Observation of the hyperfine transition in lithium-like Bismuth \ensuremath{^{209}\text{Bi}^{80+}}: Towards a test of QED in strong magnetic  fields }
\author{Matthias Lochmann}
\affiliation{Institut f\"ur Kernphysik, TU Darmstadt, Darmstadt, Germany}
\affiliation{Institut f\"ur Kernchemie, Johannes Gutenberg-Universit\"{a}t Mainz, Mainz, Germany}
\affiliation{GSI Helmholtzzentrum f\"ur Schwerionenforschung, Darmstadt, Germany}

\author{Raphael J\"{o}hren}
\affiliation{Institut f\"{u}r Kernphysik, Universit\"{a}t M\"{u}nster, M\"{u}nster, Germany}

\author{Christopher Geppert}
\affiliation{Institut f\"ur Kernphysik, TU Darmstadt, Darmstadt, Germany}
\affiliation{Helmholtz Institut Mainz, Johannes Gutenberg-Universit\"{a}t Mainz, Mainz, Germany}

\author{Zoran Andelkovic}
\affiliation{Institut f\"ur Kernchemie, Johannes Gutenberg-Universit\"{a}t Mainz, Mainz, Germany}
\affiliation{GSI Helmholtzzentrum f\"ur Schwerionenforschung, Darmstadt, Germany}

\author{Denis Anielski}
\affiliation{Institut f\"{u}r Kernphysik, Universit\"{a}t M\"{u}nster, M\"{u}nster, Germany}

\author{Benjamin Botermann}
\affiliation{Institut f\"ur Kernchemie, Johannes Gutenberg-Universit\"{a}t Mainz, Mainz, Germany}

\author{Michael Bussmann}
\affiliation{Helmholtz-Zentrum Dresden-Rossendorf, Dresden, Germany}

\author{Andreas Dax}
\affiliation{Department of Physics, University of Tokyo, Japan}

\author{Nadja Fr\"{o}mmgen}
\affiliation{Institut f\"ur Kernchemie, Johannes Gutenberg-Universit\"{a}t Mainz, Mainz, Germany}

\author{Michael Hammen}
\affiliation{Institut f\"ur Kernchemie, Johannes Gutenberg-Universit\"{a}t Mainz, Mainz, Germany}

\author{Volker Hannen}
\affiliation{Institut f\"{u}r Kernphysik, Universit\"{a}t M\"{u}nster, M\"{u}nster, Germany}

\author{Thomas K\"{u}hl}
\affiliation{GSI Helmholtzzentrum f\"ur Schwerionenforschung, Darmstadt, Germany}
\affiliation{Helmholtz-Institut Jena, Germany}

\author{Yuri A.\ Litvinov}
\affiliation{GSI Helmholtzzentrum f\"ur Schwerionenforschung, Darmstadt, Germany}

\author{Rub\'en L\'opez-Coto}
\altaffiliation[present address: ]{IFAE, Bellaterra, Spain.}
\affiliation{Institut f\"{u}r Kernphysik, Universit\"{a}t M\"{u}nster, M\"{u}nster, Germany}

\author{Thomas St\"{o}hlker}
\affiliation{GSI Helmholtzzentrum f\"ur Schwerionenforschung, Darmstadt, Germany}
\affiliation{Helmholtz-Institut Jena, Germany}
\affiliation{IOQ, Friedrich-Schiller-Universit\"{a}t Jena, Jena, Germany}

\author{Richard C. Thompson}
\affiliation{Department of Physics, Imperial College London, London, United Kingdom}

\author{Jonas Vollbrecht}
\affiliation{Institut f\"{u}r Kernphysik, Universit\"{a}t M\"{u}nster, M\"{u}nster, Germany}

\author{Andrey Volotka}
\affiliation{Institut f\"{u}r Theoretische Physik, TU Dresden, Dresden, Germany}

\author{Christian Weinheimer}
\affiliation{Institut f\"{u}r Kernphysik, Universit\"{a}t M\"{u}nster, M\"{u}nster, Germany}

\author{Weiqiang Wen}
\affiliation{Institute of Modern Physics, Lanzhou, China}

\author{Elisa Will}
\affiliation{Institut f\"ur Kernchemie, Johannes Gutenberg-Universit\"{a}t Mainz, Mainz, Germany}

\author{Danyal Winters}
\affiliation{GSI Helmholtzzentrum f\"ur Schwerionenforschung, Darmstadt, Germany}

\author{Rodolfo S\'anchez}
\affiliation{GSI Helmholtzzentrum f\"ur Schwerionenforschung, Darmstadt, Germany}
\affiliation{Helmholtz Institut Mainz, Johannes Gutenberg-Universit\"{a}t Mainz, Mainz, Germany}

\author{Wilfried N\"ortersh\"auser}
\affiliation{Institut f\"ur Kernphysik, TU Darmstadt, Darmstadt, Germany}
\affiliation{Institut f\"ur Kernchemie, Johannes Gutenberg-Universit\"{a}t Mainz, Mainz, Germany}
\affiliation{GSI Helmholtzzentrum f\"ur Schwerionenforschung, Darmstadt, Germany}

\date{\today}

\begin{abstract}
We performed a laser spectroscopic determination of the $2s$ hyperfine
splitting (HFS) of Li-like \ensuremath{^{209}\text{Bi}^{80+}} and repeated the measurement of the $1s$ HFS of H-like
\ensuremath{^{209}\text{Bi}^{82+}}.
Both ion species were subsequently stored in the Experimental Storage Ring at the GSI Helmholtzzentrum f\"ur Schwerionenforschung Darmstadt
and cooled with an electron cooler at a velocity of $\approx 0.71\,c$. Pulsed laser
excitation of the $M1$ hyperfine-transition was performed in anticollinear and
collinear geometry for \HBi and \LiBi, respectively, and observed by
fluorescence detection. We obtain $\Delta E^{(1s)}= 5086.3(11)\,\textrm{meV}$
for \HBi, different from the literature value, and $\Delta E^{(2s)}= 797.50(18)\,\textrm{meV}$ for \LiBi.
These values provide experimental evidence that a specific difference between the two splitting energies can be used to test QED calculations in the strongest static magnetic fields available in the laboratory independent of nuclear structure effects. The experimental result is in excellent agreement with the theoretical prediction and confirms the sum of the Dirac term and the relativistic interelectronic-interaction correction at a level of 0.5\% confirming the importance of accounting for the Breit interaction. 
\end{abstract}
\maketitle

Quantum electrodynamics (QED) is generally considered to be the best-tested theory in physics. In recent years a number of extremely precise experimental tests have been achieved on free particles as well as on bound states in light atomic systems. For free particles, the $g$-factor of the electron measured with ppb-accuracy \cite{Hanneke2008} constitutes the most precise test, sensitive to the highest order in $\alpha$ \cite{Aoyama2012}. In atomic systems the QED deals with the particles bound by the Coulomb field, what
makes high-precision QED calculations more complicated. The bound-state QED (BS-QED) effects in light atomic systems are expanded in parameters $Z\alpha$ and $m_{e}/M$ in addition to $\alpha$, where $Z$ is the atomic number and $m_{e}$ and $M$ are the electron and nuclear masses, respectively. The parameter $Z\alpha$ characterizes the binding strength in the Coulomb field of the nucleus, while the mass ratio $m_e/M$ is introduced for the nuclear recoil effects. Hence, tests of BS-QED are complementary to QED tests of the properties of free particles.
The investigation of H-like systems with increasing charge provides the opportunity to systematically increase the influence of the binding effect. One of the most accurate test of BS-QED on low-$Z$ ions is the measurement of the $g$-factor of a single electron bound to a Si nucleus \cite{Sturm2011}.
Entering the regime of highly charged heavy ions like Pb$^{81+}$, Bi$^{82+}$ or U$^{91+}$ the electron binding energy becomes comparable to the rest-mass energy and the parameter $Z\alpha$ can not be employed as an expansion parameter anymore.  In other words, the extremely strong electric and magnetic fields in the close surrounding of the heavy nucleus require the inclusion of the binding corrections in all orders of $Z\alpha$. Hence, BS-QED in this regime requires a very different approach and new tools to calculate the corresponding corrections, usually referred to as strong-field QED. They have been developed during recent decades \cite{Mohr1998,Sunnergren1998,Beier2000,Sapirstein2001,Yerokhin2003,Yerokhin2006,Shabaev2006,Sapirstein2008,Volotka2013a} but by far not as precisely tested as in the low-$Z$ regime. The most stringent tests are currently a Lamb shift measurement in U$^{91+}$ providing a test of QED effects on the level of 2\% \cite{Gumb05} and a measurement 
 of the $2p_{1/2} \rightarrow 2s_{1/2}$ transition energy in Li-like U$^{89+}$ that tests first and second order QED effects to a level of 0.2\% and 6\%, respectively \cite{Yerokhin2006,Beiersdorfer2005}. Here, we report on a measurement of the hyperfine splitting in heavy highly charged ions particularly sensitive to QED contributions arising from the extreme magnetic fields, which can be as strong as $10^{10}$\,T at the nuclear surface, only comparable with magnetic fields of neutron stars.
Measurements of HFS in H-like ions were performed in the past on Bi, Ho, Re, Pb and Tl \cite{Klaft1994,Crespo1996,Crespo1998,Seelig1998,Beiersdorfer2001} but did not provide conclusive tests of strong-field QED since the theoretical uncertainty arising from the insufficiently known magnetic moment distribution inside the nucleus [Bohr-Weisskopf (BW) effect] is of about the same size as the total QED contribution. 
A QED test to much higher accuracy is nevertheless possible using 
the specific difference of the HFS energies \cite{Sha01}
\begin{equation}
  \Delta ^{\prime} E = \Delta E^{(2s)}-\xi \Delta E^{(1s)}, \label{spezDiff}
\end{equation}
with $\Delta E^{(1s)}$ and $\Delta E^{(2s)}$ denoting the HFS energies of H-like \ensuremath{^{209}\text{Bi}^{82+}} and Li-like \ensuremath{^{209}\text{Bi}^{80+}}, respectively. The parameter $\xi = 0.168\,86$  
for the cancellation of the BW effect is largely model independent and the specific difference can be calculated to high accuracy \cite{Volotka2012,Andreev2012}. The theoretical value $\Delta ^{\prime} E = -61.320(6)$\,meV is dominated by the one-electron Dirac term ($-31.809$\,meV) and the interelectronic-interaction corrections of first order in $1/Z$ ($-29.995$\,meV).
Recent achievements are connected with the rigorous evaluations of the screened self-energy \cite{Volotka2009,Glazov2010} and vacuum-polarization \cite{Andreev2012} corrections, as well as with the two-photon exchange diagrams \cite{Volotka2012}. Since more than 99\% of the one-electron QED contribution cancels, the remaining QED part in the specific difference is dominated by the screened QED terms arising from the combination of the radiative and interelectronic-interaction diagrams of about 0.3\,\% ($0.193(2)$\,meV). It should be noted that the interelectronic-interaction calculated in the nonrelativistic limit yields only $\approx -9.5$\,meV \cite{Sha95} while almost 70\% are caused by relativistic effects, which can be completely (to all orders in $\alpha Z$) taken into account only within the rigorous QED approach.
Thus, investigations of the specific difference allow the many-electron QED effects in extreme electric and magnetic fields to be tested. The best case for such a measurement is $^{209}$Bi, since the transition in H-like Bi is in the UV and that of Li-like Bi still in the near infrared. The laser-spectroscopic measurement of $\Delta E^{(1s)}$ yielded 5084.0(8)\,meV already in 1993 \cite{Klaft1994}, but for $^{209}\text{Bi}^{80+}$ only a much less precise indirect x-ray emission spectroscopy measurement in an electron beam ion trap with $\Delta E^{(2s)}= 820(26)$\,meV \cite{Beiersdorfer1998} was reported.
Three attempts to measure the HFS transition in \LiBi with laser spectroscopy failed within the past 13 years even though the prediction of the transition wavelength based on the known value for the H-like Bi and the calculated $\Delta ^{\prime} E$ was expected to be very reliable. This initiated discussions about flaws in the experiment, the theoretical calculations or possible deviations from QED. Here we report the direct observation of the $M1$ hyperfine transition in the Li-like ion, unraveling this mystery and providing for the first time the experimental value for $\Delta ^{\prime} E$ to be compared with theory. This opens the perspective for studies of the HFS for tests of QED effects in a strong magnetic field of the nucleus.

\begin{figure}%
  \begin{center}
	   \includegraphics[width=0.49\textwidth]{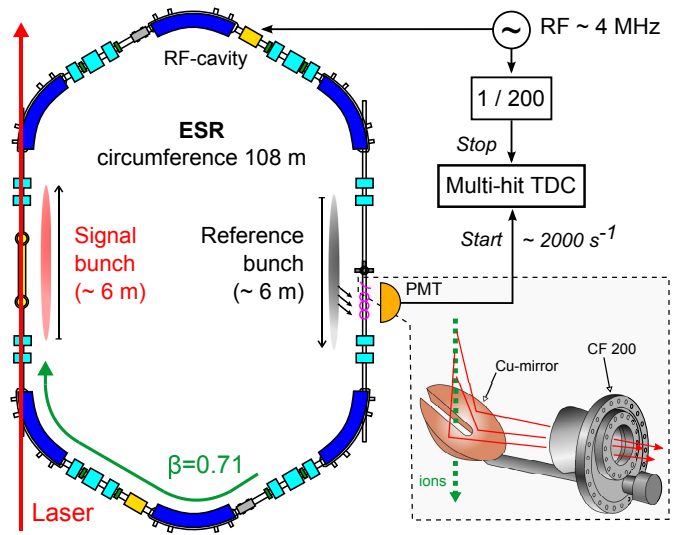}%
  \end{center}
  \caption{(color online). Experimental setup at the Experimental Storage Ring (ESR). Two ion bunches are formed by applying an rf-voltage of twice the ion's revolution frequency to an intra-ring cavity. The 'signal bunch' is repeatedly illuminated with a pulsed, blue-detuned laser for collinear excitation of Li-like \LiBi ions or a red-detuned laser for anticollinear excitation of the H-like \HBi ions (not shown). Fluorescence is detected with photomultiplier tubes (PMTs) and individual photon-arrival times relative to the rf-phase are processed in a time to digital converter (TDC). A schematic of the new detection system is also shown (see text).}%
  \label{fig:esr}%
\end{figure}

The experiment was performed at the GSI accelerator facility. First \HBi and then \LiBi ions were produced at an energy of about 400\,MeV/u and then injected into the Experimental Storage Ring (ESR) \cite{Franzke1987} (see Fig.~\ref{fig:esr}).
About 10\,s after injection, the electron-beam velocity in the electron cooler \cite{Steck2004} determines the ion velocity $\beta = \upsilon/c \approx 0.71$ at a typical ion momentum spread of $\Delta p/p\approx 10^{-4}$. The electron-cooler cathode was operated at approximately $-214$\,kV. 
The ESR orbit length is about $108.5$\,m and the ion's revolution frequency $f_\textrm{rev} \approx 2\,$MHz. A radio-frequency (rf) voltage with twice the free-revolution frequency -- measured with the Schottky analysis -- was applied to an rf cavity in the ESR, forcing the ions to circulate in two bunches of about 6\,m length each \cite{Wen2013}. One of these bunches served as a reference for residual-gas fluorescence background subtraction, whereas the other one was irradiated with the pulsed spectroscopy laser and provided signal photons on resonance.

Light at the Doppler-shifted transition wavelengths of about 590 and 640\,nm for H-like and Li-like ions, respectively, was produced by a pulsed dye laser delivering a typical pulse energy of $\approx100\,$mJ at $\lesssim10\,$ns pulse length, 30\,Hz repetition rate and $\approx18\,$GHz spectral line width.
Temporal overlap between laser pulse and ion bunch in the interaction zone inside the electron cooler was achieved by synchronizing the pump laser Q-switch signal with the bunch-generating rf-voltage.
Laser beam position and pointing at the interaction region after $\approx80\,$m transport through air were actively stabilized.

At relativistic ion velocities, fluorescence emission in the laboratory frame is neither isotropic nor monoenergetic. At $0.71c$ about $30\,\%$ of the fluorescence photons are emitted under a forward angle of $\lesssim 30^\circ$ and have wavelengths for \LiBi in the range of $640\,\textrm{nm}\lesssim\lambda_\textrm{lab}\lesssim 850\,$nm.
This fits well to the spectral sensitivity of a selected Hamamatsu R1017 photomultiplier tube (PMT) with a maximum quantum efficiency of 16\% for photons emitted in the forward direction.
Two setups for optical detection were operated during the experiment, both located in the straight section of the ESR opposite to the electron cooler (see Fig.~\ref{fig:esr}). The UV fluorescence signal of H-like ions was detected with the old mirror system, designed for measurements in H-like Pb \cite{Seelig1998}. For detection of the Li-like resonance it was also equipped with a red-sensitive PMT as in all previous attempts. Again, no resonance signal was detected on this PMT.  Instead the signal was observed with a new setup mounted in parallel, which thus proved to be the key improvement in this experiment. Its main element is an off-axis parabolic mirror with a central slit through which the ions pass \cite{Hannen2013} as it is shown in the inset of Fig.~\ref{fig:esr}. Photons emitted at angles of $1^\circ\lesssim\alpha\lesssim 20^\circ$ to the flight direction are efficiently directed to the PMT.
The photon events detected with the PMTs were processed by a multi-hit TDC with time resolution of $1/(300\,$MHz) relative to the phase of the bunching rf. Time windows for assigning detected photons to the signal bunch or the reference bunch were set and optimized offline during data analysis.

\begin{figure}[tb]
    \centering
      \includegraphics[width=0.48\textwidth]{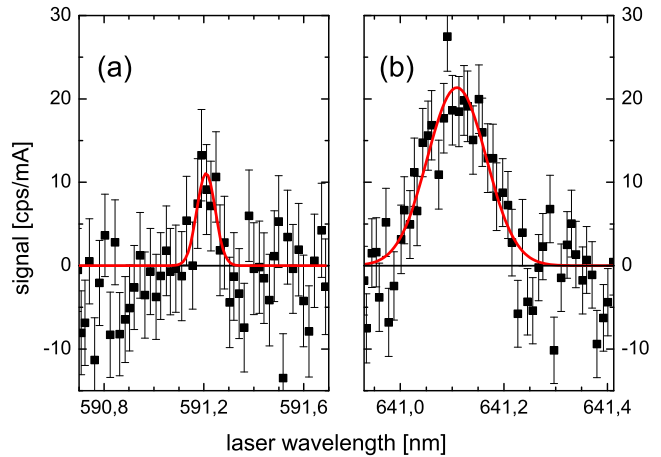}
    \caption{(Color online) Resonance of the ground-state HFS
transition in H-like \ensuremath{^{209}\text{Bi}^{82+}} (a) and Li-like \ensuremath{^{209}\text{Bi}^{80+}} (b). The signal rate is normalized to the ion current in the ESR and plotted as a function of laser wavelength. The solid line is a error-weighted nonlinear least-square fit of a Gaussian profile without background to the data.}
    \label{Resonanz}
\end{figure}

Typical background-corrected fluorescence signals obtained in a single scan for H-like (a) and Li-like ions (b), normalized to the ion current in the ESR, are shown in Fig.~\ref{Resonanz} as a function of the simultaneously recorded laser wavelength. Error bars are based on Poisson statistics.
Error-weighted fits with a Gaussian function without background yielded a linewidth (FWHM) of $\approx$40\,GHz and statistically distributed fit residuals.
Due to time constraints only seven scans could be recorded for H-like Bi. They were combined to two spectra and fitting resulted in an average $\chi_{\textrm{red}}^2=1.00$.
For Li-like \LiBi, 72 scans were performed and combined for fitting to 24 spectra with average $\chi_{\textrm{red}}^2=1.06$. The error-weighted averages of the
central laser wavelengths in the laboratory frame are
\begin{eqnarray}
	\lambda_{\textrm{lab}}^{(82+)}&=&591.183(26)\,\textrm{nm,}\label{labH}\\
	\lambda_{\textrm{lab}}^{(80+)}&=&641.112(24)\,\textrm{nm.}\label{labLi}
\end{eqnarray}
The dominant uncertainty contributions are the laser wavelength calibration (0.017\,nm) and variations or uncertainties in the ESR operating parameters causing uncertainties in the ion velocity, which were transformed to laboratory-frame wavelength uncertainties (0.018\,nm). The statistical fitting uncertainty and a possible small angle-mismatch of $\Delta \theta<2.6\,$mrad between ion beam and laser direction do not contribute significantly and were also added in quadrature.

The next and crucial step in the analysis is the transformation from the laboratory frame into the ion's rest frame, requiring the ion speed determination from the electron cooler voltage using
\begin{equation}
\beta=\sqrt{1-\gamma^{-2}}=\sqrt{1-\left(1+\frac{-eU_\textrm{e}}{m_{e}c^{2}}\right)^{-2}},
\label{betaOfU}
\end{equation}
where $e$ is the elementary charge, and $U_\textrm{e}$ is the electron accelerating potential difference.

Starting from the electron-cooler set voltages during the spectroscopy of \HBi ($-213.900\,$kV) and \LiBi ($-213.890\,$kV), we took into account several corrections and uncertainties, based on several test measurements before, during and after the beamtime, which will be described in a forthcoming paper.
We obtained the calculated effective acceleration voltages $U_\textrm{e}=-214.00(11)\,\textrm{kV}$ and $U_\textrm{e}=-213.93(11)\,\textrm{kV}$ for H-like and Li-like ions, respectively. The relative uncertainty of approximately $5 \times 10^{-4}$ is comparable to that estimated in \cite{Klaft1994} for the previous measurement of H-like bismuth, but at a considerably higher velocity.  Subsequent attempts to improve our voltage calibration were hampered by technical defects and resulting major repairs both in the voltage supply and in the voltmeter shortly after the measurements. The rest frame transition wavelengths are calculated using the relativistic Doppler formula $\lambda_0=\lambda_{\textrm{lab}} \gamma \left(1\mp\beta \right)$ for \HBi and \LiBi, respectively. The results are summarized in Table \ref{tab:ergebnisse}. We keep the uncertainty due to the voltage calibration (first parentheses) separated from the other uncertainties, because they are strongly correlated for the measurements of H-like and Li-like bismuth (we expect the electron-cooler calibration to be unchanged during the consecutive spectroscopy of both charge states), since the spectroscopy on one species is performed in collinear and on the other one in anticollinear geometry the correlation actually increases the uncertainty. If -- due to a miscalibration -- the electron-cooler voltage assumed in the analysis is smaller than the one actually used in the experiment, the calculated H-like HFS is too small, while the extracted HFS in the Li-like ion is too large, and vice versa.
We obtained the calculated effective acceleration voltages $U_\textrm{e}=-214.00(11)\,\textrm{kV}$ and $U_\textrm{e}=-213.93(11)\,\textrm{kV}$ for H-like and Li-like ions, respectively. The relative uncertainty of approximately $5 \times 10^{-4}$ is comparable to that estimated in \cite{Klaft1994} for the previous measurement of H-like bismuth, but at a considerably higher velocity.  Subsequent attempts to improve our voltage calibration were hampered by technical defects and resulting major repairs both in the voltage supply and in the voltmeter shortly after the measurements. The rest frame transition wavelengths are calculated using the relativistic Doppler formula $\lambda_0=\lambda_{\textrm{lab}} \gamma \left(1\mp\beta \right)$ for \HBi and \LiBi, respectively. The results are summarized in Table \ref{tab:ergebnisse}. We keep the uncertainty due to the voltage calibration (first parentheses) separated from the other uncertainties, because they are strongly correlated for the measurements of H-like and Li-like bismuth (we expect the electron-cooler calibration to be unchanged during the consecutive spectroscopy of both charge states), since the spectroscopy on one species is performed in collinear and on the other one in anticollinear geometry the correlation actually increases the uncertainty. If -- due to a miscalibration -- the electron-cooler voltage assumed in the analysis is smaller than the one actually used in the experiment, the calculated H-like HFS is too small, while the extracted HFS in the Li-like ion is too large, and vice versa.
\begin{table}
	\caption{\label{tab:ergebnisse}Experimental and theoretical values for the rest frame wavelengths $\lambda_0$ and transition energies $\Delta E$ of the HFS-transitions in highly charged $^{209}$Bi.
	For wavelengths and energies, the first (correlated) uncertainty arises from the dominant voltage calibration uncertainty, and the second one arises in about equal parts from other voltage uncertainty contributions and the laboratory frame wavelength uncertainties in Eqs.~(\ref{labH} and \ref{labLi}). For literature values
	only the total uncertainty is given. 
}
	\centering
	\begin{tabular}{llll}
		\hline\hline
		 Ref. & $eU_\textrm{e}$(keV) & $\lambda_0$(nm) & $\Delta E$(meV)\\
		\hline
		\HBi  \\
		 this &--214.00(11)             & \phantom{1}243.76(5)(2)           & 5086.3(11)(03)\\
		 exp \cite{Klaft1994}       & --120.00(6)  & \phantom{1}243.87(4) & 5084.0(8)\\
                 theo \cite{Shabaev1997}    &   & \phantom{1}243.0(13) & 5101(27)\\
		 theo \cite{Tomaselli1998}  &   & \phantom{1}245.13(58)& 5058(12)\\
		 theo \cite{Senkov2002}     &   &                        & 5111(-6/+21) \\
		\hline
		\LiBi \\
		 this          &--213.93(11)    & 1554.66(33)(10)          & \phantom{1}797.50(17)(05) \\
		 exp \cite{Beiersdorfer1998}&   & 1512(50)& \phantom{1}820(26) \\
		 theo \cite{Shabaev2000}    &   & 1555.4(4)& \phantom{1}797.1(2) \\
                 theo \cite{Boucard2000}    &   & 1563.9 &\phantom{1}792.8 \\
                 theo \cite{Sapirstein2001} &   & 1555.3(3)&\phantom{1}797.15(13) \\
		 theo \cite{Volotka2012}    &   & 1555.3(3)\phantom{63(10)}&\phantom{1}797.16(14) \\
		\hline
		\hline
	\end{tabular}
\end{table}
In the case of \HBi both experimental values obtained at the ESR are in agreement with a theoretical result for which the BW effect was evaluated within the single-particle nuclear model ~\cite{Shabaev1997} but its uncertainty fully masks the QED effects. More elaborated calculations of the BW effect employing many-particle nuclear models \cite{Tomaselli1998,Senkov2002} disagree with the experimental values. In the case of \LiBi all theoretical predictions listed in Tab.\,\ref{tab:ergebnisse} were obtained by extracting the BW-correction from the experimental result for \HBi \cite{Klaft1994}. The discrepancy of $\approx 0.4$~meV between our value for $\Delta E(\LiBi)$ and the ones predicted in \cite{Shabaev2000,Sapirstein2001,Volotka2012} can be directly traced back to the difference between the two measurements of \HBi ($2.3$\,meV).
This is also reflected by the excellent agreement between $\Delta' E = -61.37(35)(08)\,$meV determined from the HFS reported here and the theoretical prediction $\Delta' E = -61.320(6)\,$meV \cite{Volotka2012,Andreev2012} which is free from experimental input. 
This also confirms the dominating contributions in the specific difference, the Dirac term and the interelectronic-interaction corrections, on a $5\times10^{-3}$ level. The latter -- being up to $70\,\%$ of relativistic origin -- can only be evaluated within the rigorous QED approach. Thus, we have effectively tested the relativistic interelectronic-interaction in presence of a strong magnetic field and unambiguously confirmed the importance of accounting for the Breit interaction.

Our value for the HFS in \LiBi is two orders of magnitude more precise than the only experimental value reported so far, determined indirectly via x-ray emission spectroscopy \cite{Beiersdorfer1998}. For H-like \HBi the extracted HFS is on a $2\sigma$ level inconsistent with the one reported by Klaft and coworkers \cite{Klaft1994}, although obtained with a similar experimental setup at GSI. If we suppose Klaft \textit{et al}.'s value is correct, the most likely reason for the discrepancy is a miscalibration of our high-voltage measurement, leading to a deviation $\delta U_\textrm{mis}$ between the real effective voltage and the calculated effective voltage. To obtain a result for the rest-frame frequency of Li-like \LiBi independent from the voltage calibration, we use the relation
\begin{equation}
\lambda_{\textrm{lab}}^{(82+)} \lambda_{\textrm{lab}}^{(80+)} = \lambda_0^{(80+)} \lambda_0^{(82+)}\label{product}
\end{equation}
which must be fulfilled, because Bi$^{82+}$ was measured anticollinearly and Bi$^{80+}$ collinearly. The velocity dependence is completely removed under the premise that the measurement of both ionic states were performed at the same ion velocity. In the experiment a small difference occured, corresponding to an electron-cooler voltage difference of only 69\,V that was taken into account as a laboratory wavelength shift of $-0.086$\,nm, being sufficiently independent on the absolute cooler voltage.
By solving (\ref{product}) for $\lambda_0^{(80+)}$ and calculating the specific difference based on this value and $\lambda_0^{(82+)}$ from Ref. \cite{Klaft1994} we obtain $\Delta' E = -60.63(19)$\,meV, which disagrees with the theoretical prediction on the $>3\sigma$ level. This and the excellent agreement of our voltage-based analysis with theory supports our result, but calls for more measurements at higher precision. The uncertainty stated by Klaft \textit{et al}. \cite{Klaft1994} is clearly dominated by the voltage uncertainty, which might have been underestimated.

In summary, we have remeasured the HFS in \HBi and directly observed and measured the HFS-transition in \LiBi, improving accuracy hundred times compared to a previous indirect measurement. We found that previous failures to observe this transition were most likely caused by insufficient sensitivity of the optical detection system. Our system increased the detection efficiency especially for the blue-shifted photons in the forward cone and is an important development for laser spectroscopy on highly relativistic ion beams in storage rings.
The experimental uncertainty is dominated by the electron cooler voltage calibration uncertainty. Our results confirm the calculated $\Delta'E$ on a level of $5\times 10^{-3}$ but are not sufficiently accurate to test the QED contributions.

Using a more accurate HV measurement of the electron-cooler voltage will improve the measurement accuracy by at least one order of magnitude. This will allow for a more precise determination of $\Delta^{\prime}E$ and the QED test in this strong magnetic field regime.
The next step will be trap-assisted laser spectroscopy on cooled \HBi and \LiBi in the SPECTRAP Penning trap currently being commissioned at GSI \cite{Andelkovic2013}, promising relative accuracies three orders of magnitudes better for both charge states. This will test the QED contribution on a level of a few percent but requires the transition wavelength to be known with at least the accuracy provided here in order to find the weak and narrow transition by fluorescence spectroscopy with a cw laser.

\begin{acknowledgments}
This work has been supported by BMBF under Contracts No. 06MS9152I and No. 05P12RDFA4, from the Helmholtz Association under Contract No. VH-NG-148, and from the Helmholtz International Center for FAIR (HIC for FAIR) within the LOEWE program by the state of Hesse.
T.S., Y.A.L., and W.W. acknowledge support from the Helmholtz-CAS Joint Research Group HCJRG-108,
and W.W. and Y.A.L received support from the BMBF grant in the framework of the Internatonale Zusammenarbeit in Bildung und Forschung Project No. 01DO12012. M. L. acknowledges support from HGS-Hire.
Helpful discussions with V.~Shabaev and continuous support from H.-J.~Kluge are acknowledged. Support from the accelerator and storage ring divisions at GSI is gratefully acknowledged. We thank S. Minami and N. Kurz at the GSI experiment electronics department for their support in the data acquisition and the VUPROM development. We thank Spectra Physics and TEM Messtechnik for their support and Sirah Lasertechnik for outstanding customer support during our beamtime.
\end{acknowledgments}


\end{document}